\documentclass{aastex63}

\usepackage[mathlines]{lineno}
\usepackage{url}

\received{XXX X, 202X}
\revised{XXX X, 202?}
\accepted{\today}
\submitjournal{ApJ}

\graphicspath{{./}{plots/}}

\begin{document}

\title{ANTARES search for point-sources of neutrinos using astrophysical catalogs: a likelihood stacking analysis}

\correspondingauthor{J. Aublin}
\email{julien.aublin@apc.in2p3.fr}

\author{A.~Albert}
\affiliation{\scriptsize{Universit\'e de Strasbourg, CNRS,  IPHC UMR 7178, F-67000 Strasbourg, France}}
\affiliation{\scriptsize Universit\'e de Haute Alsace, F-68200 Mulhouse, France}

\author{M.~Andr\'e}
\affiliation{\scriptsize{Technical University of Catalonia, Laboratory of Applied Bioacoustics, Rambla Exposici\'o, 08800 Vilanova i la Geltr\'u, Barcelona, Spain}}

\author{M.~Anghinolfi}
\affiliation{\scriptsize{INFN - Sezione di Genova, Via Dodecaneso 33, 16146 Genova, Italy}}

\author{G.~Anton}
\affiliation{\scriptsize{Friedrich-Alexander-Universit\"at Erlangen-N\"urnberg, Erlangen Centre for Astroparticle Physics, Erwin-Rommel-Str. 1, 91058 Erlangen, Germany}}

\author{M.~Ardid}
\affiliation{\scriptsize{Institut d'Investigaci\'o per a la Gesti\'o Integrada de les Zones Costaneres (IGIC) - Universitat Polit\`ecnica de Val\`encia. C/  Paranimf 1, 46730 Gandia, Spain}}

\author{J.-J.~Aubert}
\affiliation{\scriptsize{Aix Marseille Univ, CNRS/IN2P3, CPPM, Marseille, France}}

\author{J.~Aublin}
\affiliation{\scriptsize{Universit\'e de Paris, CNRS, Astroparticule et Cosmologie, F-75006 Paris, France}}

\author{B.~Baret}
\affiliation{\scriptsize{Universit\'e de Paris, CNRS, Astroparticule et Cosmologie, F-75006 Paris, France}}

\author{S.~Basa}
\affiliation{\scriptsize{Aix Marseille Univ, CNRS, CNES, LAM, Marseille, France }}

\author{B.~Belhorma}
\affiliation{\scriptsize{National Center for Energy Sciences and Nuclear Techniques, B.P.1382, R. P.10001 Rabat, Morocco}}

% ...

\collaboration{138}{(ANTARES Collaboration)}

\author{S. Buson}
\affiliation{\scriptsize{Universität Würzburg, Institut für Theoretische Physik und Astrophysik, Campus Hubland Nord, Emil-Fischer-Str. 31
97074 Würzburg, Germany}}




\begin{abstract}
A search for astrophysical point-like neutrino sources using the data collected by the ANTARES detector between January 29, 2007 and December 31, 2017 is presented. A likelihood stacking method is used to assess the significance of an excess of muon neutrinos inducing track-like events in correlation with the location of a list of possible sources. Different sets of objects are tested in the analysis: a) a sub-sample of the \textit{Fermi} 3LAC catalog of blazars, b) a jet-obscured AGN population, c) a sample of soft gamma-ray selected  radio galaxies, d)  a star-forming galaxy catalog , and e) a public sample of 56 very-high-energy track events from the IceCube experiment.

None of the tested sources shows a significant association with the sample of neutrinos detected by ANTARES. The smallest p-value is obtained for the radio galaxies catalog with an equal weights hypothesis, with a pre-trial p-value equivalent to a $2.8 \, \sigma$ excess, equivalent to $1.6 \, \sigma$ post-trial.

In addition, the results of a dedicated analysis for the blazar MG3 J225517+2409 are also reported: this source is found to be the most significant within the \textit{Fermi} 3LAC sample, with 5 ANTARES events located at less than one degree from the source. This blazar showed evidence of flaring activity in \textit{Fermi} data, in space-time coincidence with a high-energy track detected by IceCube. An \emph{a posteriori} significance of $2.0\, \sigma$ for the combination of ANTARES and IceCube data is reported.

\end{abstract}



\section{\label{sec:level1}Introduction}

The origin and nature of cosmic rays at very high energy is a long-standing puzzle that is still not completely resolved after decades of theoretical and experimental efforts. The fact that cosmic rays are mostly charged particles, experiencing magnetic deflections during their propagation, makes the identification of their sources very difficult. 

In contrast, neutrinos ($\nu$) are neutral particles that can travel without absorption nor deflection from their source when reaching the Earth. Neutrinos of cosmic origin are expected to be produced via the decay of charged pions and kaons, generated in hadronic interactions of cosmic rays with gas or radiation in their acceleration sites or during their propagation. 
The decay of neutral pions also produces gamma-rays ($\gamma$) that should be detectable, thus intimately connecting these three messengers: cosmic rays, neutrinos and gamma-rays.

The observation of a diffuse astrophysical neutrino flux has been established in 2013 by the IceCube Collaboration with a high level of significance \cite{ICAstroFlux}, with the detection of a excess of neutrinos above the expected background. Their spatial distribution compatible with isotropy favors an extra-galactic origin.

Moreover, the first association between a high-energy neutrino and a cosmic source has been reported by IceCube \cite{TXS_MMessengerObs} in September 2017, when a through-going muon track with a deposited energy of $\sim24$ TeV has been detected in spatial coincidence with the position of the blazar TXS 0506+056. The high-energy neutrino occurred during a period of intense gamma-ray emission observed by \textit{Fermi}-LAT and MAGIC telescopes \cite{TXS_MMessengerObs}. A subsequent study \cite{TXSFollow} based on IceCube’s archival data in the direction of the blazar has reported the presence of a $3.5 \,\sigma$ candidate TeV-neutrino flare between September 2014 and March 2015 positionally consistent with TXS 0506+056. ANTARES data do not show strong evidence of neutrinos in the direction of this blazar \cite{TXS_ANTARES}.

However, during the IceCube neutrino flare period, the gamma-ray flux of the source was one order of magnitude lower than its value during the intense gamma-ray activity period and showed no sign of time variability.
If the neutrino signal observed in the direction of TXS 0506+056 is truly of astrophysical origin and related to the blazar, this implies that the relation between the observed gamma-ray flux and the high-energy neutrino flux is not simple, and can depend on the individual properties of each source. More multi-messenger observations are then needed to constrain the models and gather evidence to identify the sources of high-energy neutrinos.\\

The recent findings using IceCube data (\cite{Padovani2016}, \cite{Investig2Blazars}, \cite{DissectBlazars}), support the idea that blazars could produce an observable high-energy neutrino flux.
Among the various potential sources of high-energy neutrinos, active galactic nuclei (AGN) and more specifically blazars are therefore of particular interest. Indeed, when one of the relativistic jets emitted by the central black hole is pointing close to the Earth line of sight, the relativistic Doppler effect can produce a boosted flux of electromagnetic radiation and potentially related neutrinos, if hadronic interactions occur within the jet. The mechanism of production of high-energy neutrinos from blazars has been studied in numerous papers \cite{Mannheim1995}, \cite{Halzen1997}, \cite{Mucke2003}, \cite{Murase2014}, \cite{Tavecchio2015}, and is still an active field of investigation.

From the experimental point of view, different methods have been used to search for neutrino point-sources on the full observable sky \cite{PSpaper}, \cite{CombinedPSIC_ANTARES}, \cite{IC_10yrsPS}, including searches over a predefined target list \cite{ICBlazars}, \cite{ICBlazars_ICRC17}, \cite{ICBlazars_ICRC19}.
The large number of trials associated with the full sky searches implies that only very bright objects can be detected with a $>5\, \sigma$ significance with such methods. Searching among a limited set of directions, for example from a catalog of sources, is an efficient way to reduce the trial factors and also allows population studies to be performed. 

The present analysis uses a likelihood stacking method, allowing the significance of an excess of neutrino events in the direction of selected sample of sources to be assessed. The potential neutrino emitters are considered together, as a common population from which neutrinos could be produced. The ANTARES data provides an independent measurement with respect to the IceCube data, offering a complementary sensitivity in the Southern Hemisphere (see figure 1 in \cite{CombinedPSIC_ANTARES}).

The ANTARES data collected between 2007 and 2018 is used to find possible correlations with several catalogs. The \textit{Fermi} 3LAC catalog \cite{Fermi3LAC} is used to test the blazar origin hypothesis; other catalogs of astrophysical objects are also considered: dust-obscured AGN \cite{ObscuredAGN},  very bright radio galaxies \cite{Radiogals}, and star-forming galaxies \cite{SFGPaper}.
As an additional and independent test, the correlation of ANTARES neutrinos with 51 track events from the IceCube public high-energy neutrino candidates \cite{IC_Spectrum}, \cite{IC_HESE6} is also evaluated.\\

The paper is organized as follows: the data set and the method are presented in sections \ref{sec:level2} and \ref{sec:level3}, the target catalogs are described in section \ref{sec:level4} and the results are shown in section \ref{sec:level5}. An investigation of the most interesting individual objects is presented in subsection \ref{sec:subsec1}, and a dedicated analysis of the blazar MG3 J225517+2409 is performed in subsection \ref{sec:subsec2}. Finally, the results are summarized and discussed in section \ref{sec:level6}.

\section{\label{sec:level2}Data set}

The ANTARES neutrino telescope is a water Cherenkov detector operating since 2007 on the bottom of the Mediterranean Sea, 40-km off-shore Toulon (France), that detects high-energy neutrinos by observing the passage in water of relativistic particles produced by interactions of neutrinos in the vicinity of the instrumented volume \cite{ANTARES_NIM}.\\ 

The data sample used in the present analysis is presented in \cite{PSpaper} and consists of 8754 track-like events detected by ANTARES between January 29, 2007 and December 31, 2017, for an overall livetime of 3125.4~days. These events are almost 100\% induced by charged current interactions of $\nu_\mu$, the detected track corresponding to the path of the emitted muon.

The search for point-like sources requires a very good angular resolution. The event selection has been optimized to detect a point-like source with an energy spectrum $\propto E^{-2}$. The present data selection uses the same cuts that were defined in the 9 yr search \cite{PSpaper}, applied to an extended data set that includes two additional years.
As the in-depth description of the selection criteria and data/Monte Carlo comparison can be found in \cite{PSpaper}, only the most relevant parameters are described in the following.\\ 

The candidate neutrinos are selected by first considering mainly up-going directions, where the reconstructed zenith angle $\theta$ is reduced to values $\cos \theta > -0.1$, and by further cutting on the estimated angular error $\beta$ and the quality parameter $\Lambda$ of the likelihood reconstruction. 
The atmospheric muons, which constitute the dominant contribution to the background, are mostly suppressed by the cut on the quality parameter $\Lambda$ ($\sim 4$ orders of magnitude), and by the zenith angle cut ($\sim 1$ additional order of magnitude). 

After the selection procedure, a residual contamination of $\sim 13$\% is estimated \cite{PSpaper} to come from mis-reconstructed atmospheric muons. In comparison, about $\sim 26$ signal events would be expected in total from a reference astrophysical diffuse $E^{-2}$ flux of $\nu_{\mu} + \bar{\nu}_{\mu}$ with a normalization of $10^{-8}$ GeV$^{-1}$cm$^{-2}$s$^{-1}$ at $1\,$GeV.   

The energy estimator used in the analysis is based on the measure of the energy deposit of the muon in water \cite{dEdX}. The selected tracks have estimated neutrino energies between $\sim 100$ GeV and $\sim 1 $ PeV,  with a median angular resolution better than $0.4^\circ$ above 10 TeV.

\section{\label{sec:level3}Description of the method}

\subsection{Test statistic}

An extended maximum likelihood method \cite{ExtendedMaxLL} is used to reject the null hypothesis $H_0$ where only background is assumed to be present, from an alternative signal plus background hypothesis $H_1$. 
The log-likelihood for both hypotheses $H_0$ and $H_1$ is written as respectively:

\begin{eqnarray}
 \ln \mathcal{L}(x | \mathrm{H}_0) &&=  \sum_i^N \ln \left[\mu_b B(x_i) \right] - \mu_b \\
 \ln \mathcal{L}( x | \mathrm{H}_1)&& = \sum_i^N \ln \left[ \mu_s S(x_i) + \mu_b B(x_i) \right] - \mu_s - \mu_b \nonumber ,
\end{eqnarray}
where $N$ is the total number of observed neutrino candidates, $S(x_i)$ is the probability density function (PDF) of the signal and $B(x_i)$ the PDF of the background. 
The free parameters are the estimated number of signal $\mu_s$ and background $\mu_b$ events.
The test statistic is then defined as:
\begin{equation}
    TS=\ln  \left ( \frac{ \max ( \mathcal{L}(x  | \mathrm{H}_1))  }{ \max (\mathcal{L}(x | \mathrm{H}_0) )} \right).
\end{equation}

\subsection{Probability density function of signal and background}
\subsubsection{Signal}
For a given catalog, the signal term is proportional to the expected number of neutrinos that should be detected as track-like events by ANTARES, from the directions of the $N_\mathrm{sources}$ objects, each one of them emitting a  $\nu_{\mu} + \bar{\nu}_{\mu}$ neutrino flux: $\Phi_j(E)=\Phi^0_j \left( \frac{E}{1 \mathrm{GeV}} \right)^{-\gamma}$, the flux normalization $\Phi_j^0$ (GeV$^{-1}$cm$^{-2}$s$^{-1}$) being different for each source $j$. The spectral index $\gamma$ is assumed to be $\gamma=2.0$ in the analysis, and is fixed in the likelihood maximization. 

The signal probability density function is then written as:
\begin{equation}
   S(x_i)= \frac{1}{\sum w_j} \sum_{j=1}^{N_\mathrm{sources}} w_j \,s_j(x_i),
\end{equation}
where $s_j(x_i)$ is the contribution of the $j^\mathrm{th}$ object, with weight $w_j$. 
The weight of the $j^\mathrm{th}$ source is defined as: 
\begin{equation}
w_j= w_j^\mathrm{model} \, \mathcal{A}(\delta_j),
\end{equation}
where $\mathcal{A}(\delta)$ is the declination-dependent acceptance of the neutrino track sample computed for an $E^{-2}$ energy spectrum. 

The function $\mathcal{A}(\delta)$ allows the computation of the number of signal events expected for a single point-source having a flux normalization $\Phi_0$ simply via the relation: \begin{equation}
n(\delta)=\Phi_0 \,\mathcal{A}(\delta).
\end{equation}

For a given energy spectrum $\Phi(E)=\Phi_0 \left( \frac{E}{1 \mathrm{GeV}} \right)^{-\gamma}$, the acceptance is computed from the effective area \cite{PSpaper} of the detector  $\mathcal{A}_{\mathrm{eff}}(E,\delta, t)$ with the relation:
\begin{equation}
\mathcal{A}(\delta)= \int \int \mathrm{d}t \mathrm{d}E \, \mathcal{A}_{\mathrm{eff}}(E,\delta,t) \, \left( \frac{E}{1 \mathrm{GeV}} \right)^{-\gamma} ,
\end{equation}
where the integral is performed on a sufficiently large energy range $[100$ GeV$-10$ PeV], and for the total livetime considered in the analysis (3125.4 days). The acceptance obtained for a pure $E^{-2}$ energy spectrum  is shown on figure~5 of \cite{PSpaper}.\\

The weighting scheme aims to set the weights to be proportional to the true neutrino flux. Since this is unknow, two different weighting assumptions are considered:
\begin{itemize}
    \item $w_j^\mathrm{model}=\Phi_j^0$ 
    \item $w_j^\mathrm{model}=1$ (equal flux assumption),
\end{itemize}
where $\Phi_j^0$ is the photon flux observed in a relevant part of the electromagnetic spectrum, to be used as a proxy for the neutrino flux.
As the relation between the flux emitted in a given band of the electromagnetic spectrum and the neutrino flux is rather uncertain, the equal flux assumption ($w_j^\mathrm{model}=1$) provides an additional model-independent test, which could be more sensitive if the weights in the likelihood are very different from the true (unknown) ones.

 The individual source contribution to the signal for the $i^{\rm th}$ event is:
\begin{equation}
    s_j(x_i)= \mathcal{F}(\alpha_{ij}, E_i, \beta_i),  
\end{equation}
where $\mathcal{F}$ is the energy dependent point-spread function, defined as the probability density for the reconstructed event direction to lie within an angular distance $\alpha_{ij}$ from the true source direction. 

\begin{figure}[htbp]\centering
\includegraphics[width=0.5\linewidth]{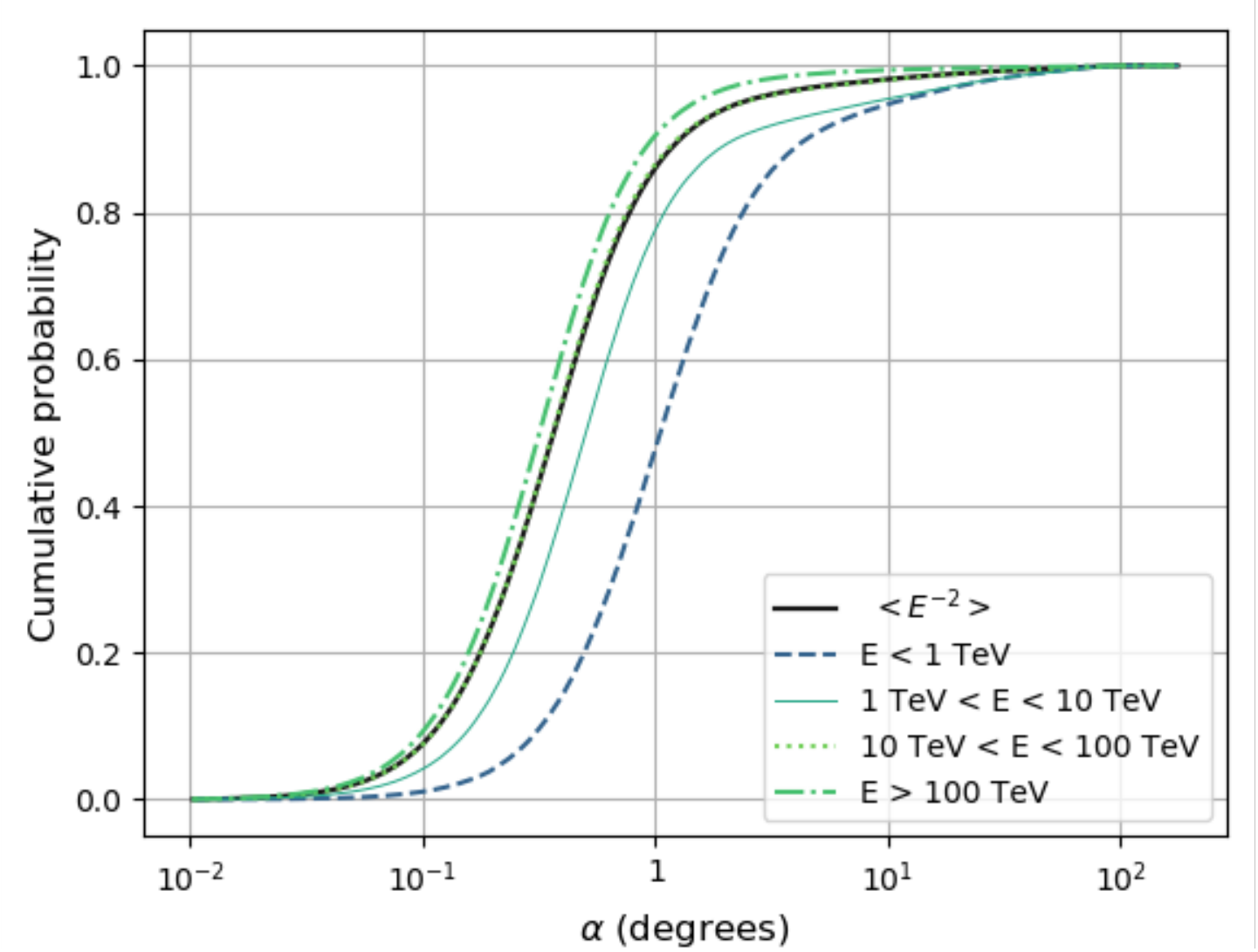}
\caption{Cumulative probability to reconstruct a track event direction within an angle  $\alpha$ with respect to the true Monte Carlo neutrino direction. The thick solid black line corresponds to the average value for an $E^{-2}$ energy spectrum, while the thinner lines are obtained for different energy ranges (still with $E^{-2}$ weights included).}
\label{fig:psf}
\end{figure}

Figure \ref{fig:psf} shows that on average, for an $E^{-2}$  spectrum, 50\% of the track events in the point-source sample are reconstructed within an angle of $0.4^\circ$ of the true neutrino direction.

The most relevant parameters that determine the shape of the point-spread function are the estimated energy $E_i$ and the angular reconstruction uncertainty $\beta_i$ coming from the track reconstruction procedure.

The function $\mathcal{F}$ is evaluated using a collection of 3D histograms built from Monte Carlo simulations, in declination bands of $\Delta \sin\delta=0.2$. For a given track event with declination $\delta$, energy $E$ and angular uncertainty $\beta$, the histogram corresponding to the correct declination band is used, and the value of the signal is obtained by linear interpolation in the $(\alpha,E,\beta)$ space.\\
 

\subsubsection{Background}
\begin{figure}[htbp]\centering
\includegraphics[width=0.48\linewidth]{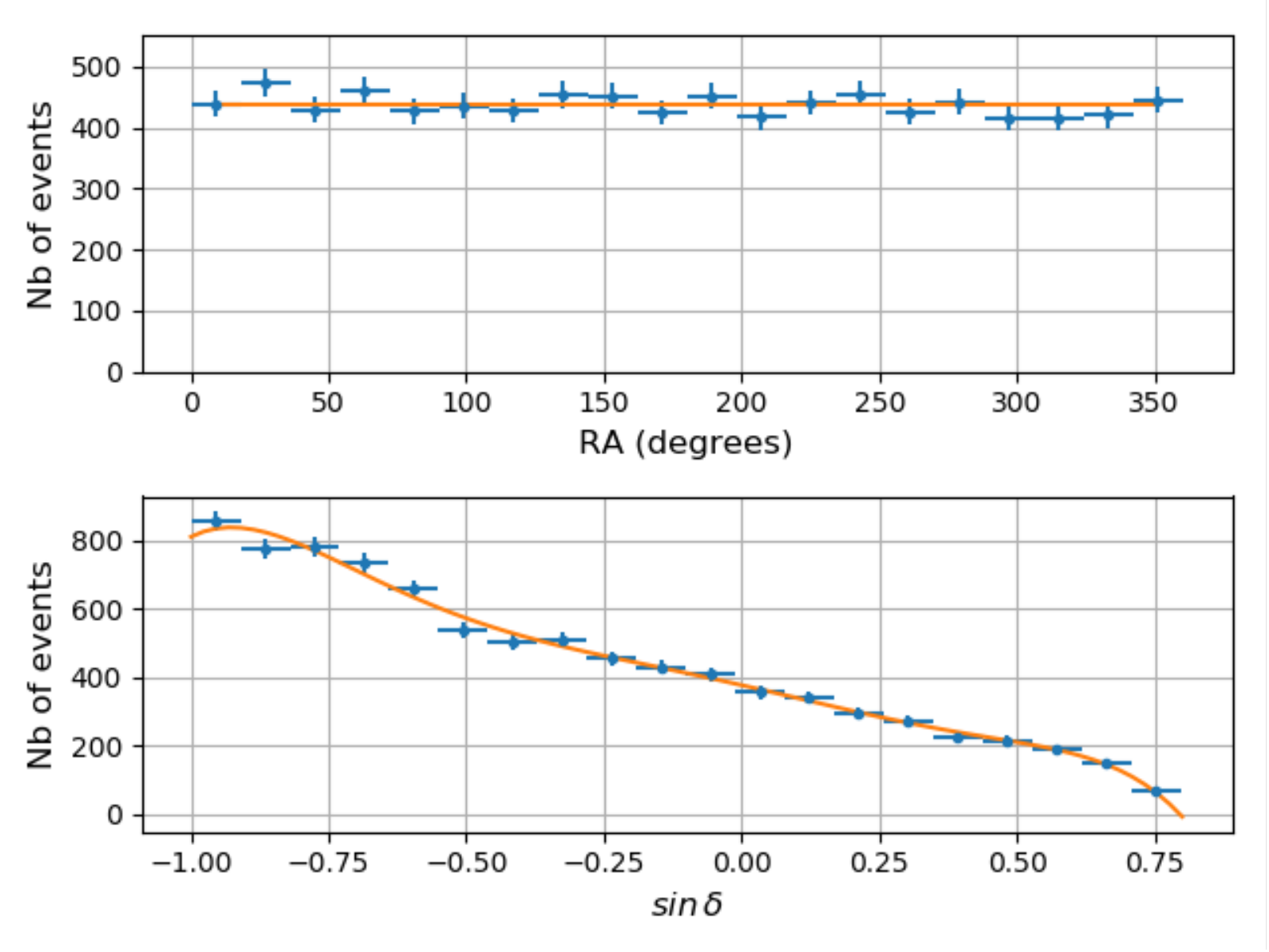}\hfill \includegraphics[width=0.48\linewidth]{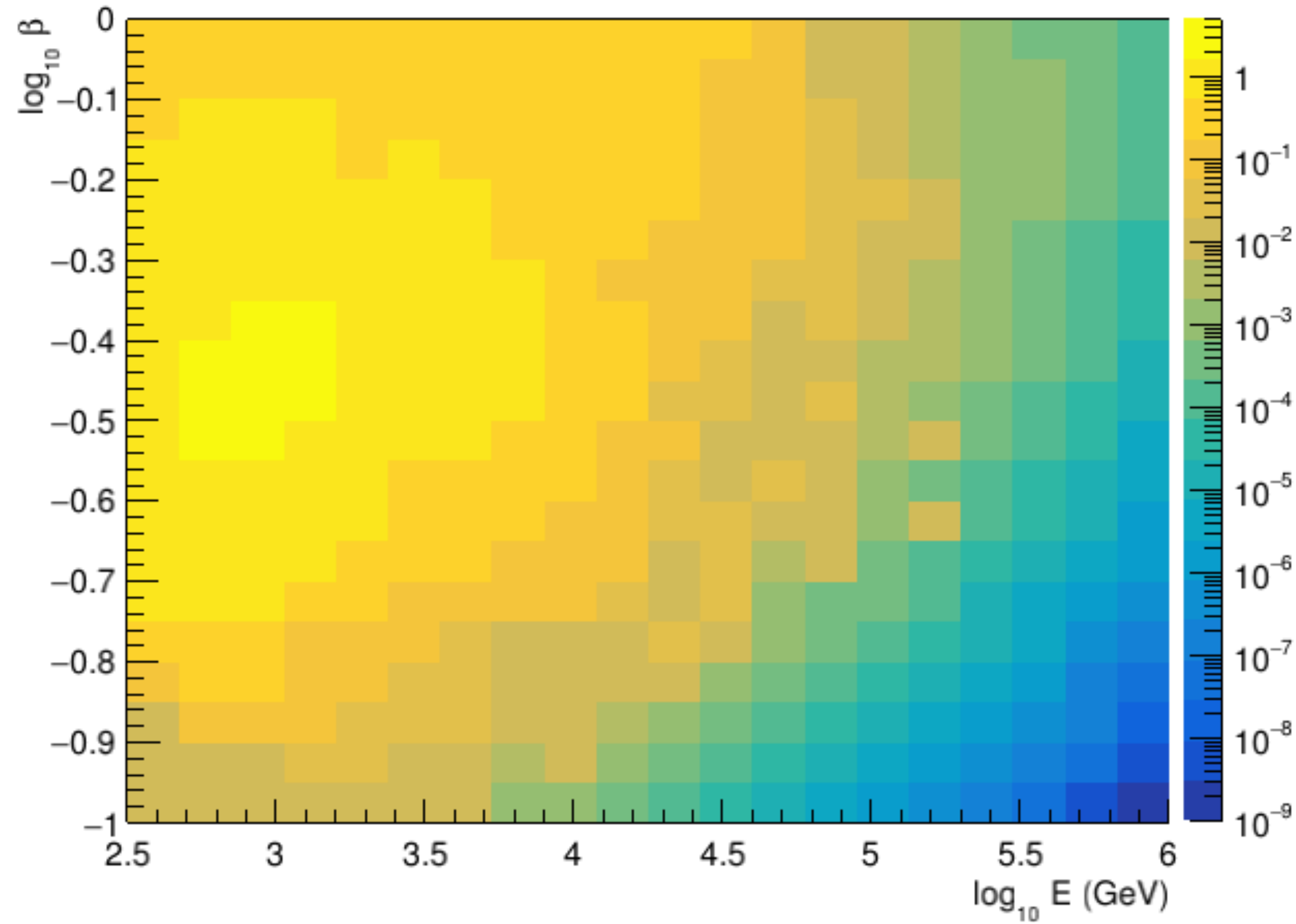}

\caption{Distribution of the right ascension (RA) and of the $\sin \delta$ (\textit{left} plots) obtained for events passing the analysis cuts in the data collected by ANTARES between January 2007 and December 2017. The solid line represents the polynomial fit function used in the likelihood evaluation. The two-dimensional histogram representing the function $f _\mathrm{b}(E,\beta)$ in color code is shown on the right plot (arbitrary units).}
\label{fig:bkg}
\end{figure}

A conservative approach is used to determine the background term. As the contribution of an astrophysical signal is expected to be small, real data are used to build the background PDF. The limited statistics available in real data compared to Monte Carlo do not allow the construction of a 3 dimensional histogram, therefore the expression of the PDF is factorized as:

\begin{equation}
  B(x_i) =\mathcal{B}(\sin \delta_i) \, f_\mathrm{b}(E_i,\beta_i) ,
\end{equation}
where $\mathcal{B}(\sin \delta)$ is the observed distribution of the sine of the declination, and $f_b$ the distribution of energy $E$ and angular uncertainty $\beta$ derived in data. The distribution of ANTARES track events measured during more than ten years is independent of the right ascension (see figure \ref{fig:bkg} top left), because of the Earth’s rotation averaging the non-uniformity of the detector's exposure in local coordinates.

 Figure \ref{fig:bkg} (bottom left) shows the rate of events as a function of $\sin \delta$, fitted by a polynomial function which is used in the likelihood computation. The value of the function $f _\mathrm{b}(E,\beta) $ is evaluated by linear interpolation in the 2D histogram shown in figure \ref{fig:bkg} (right). 
 

\begin{figure*}[t]\centering
\includegraphics[width=18cm]{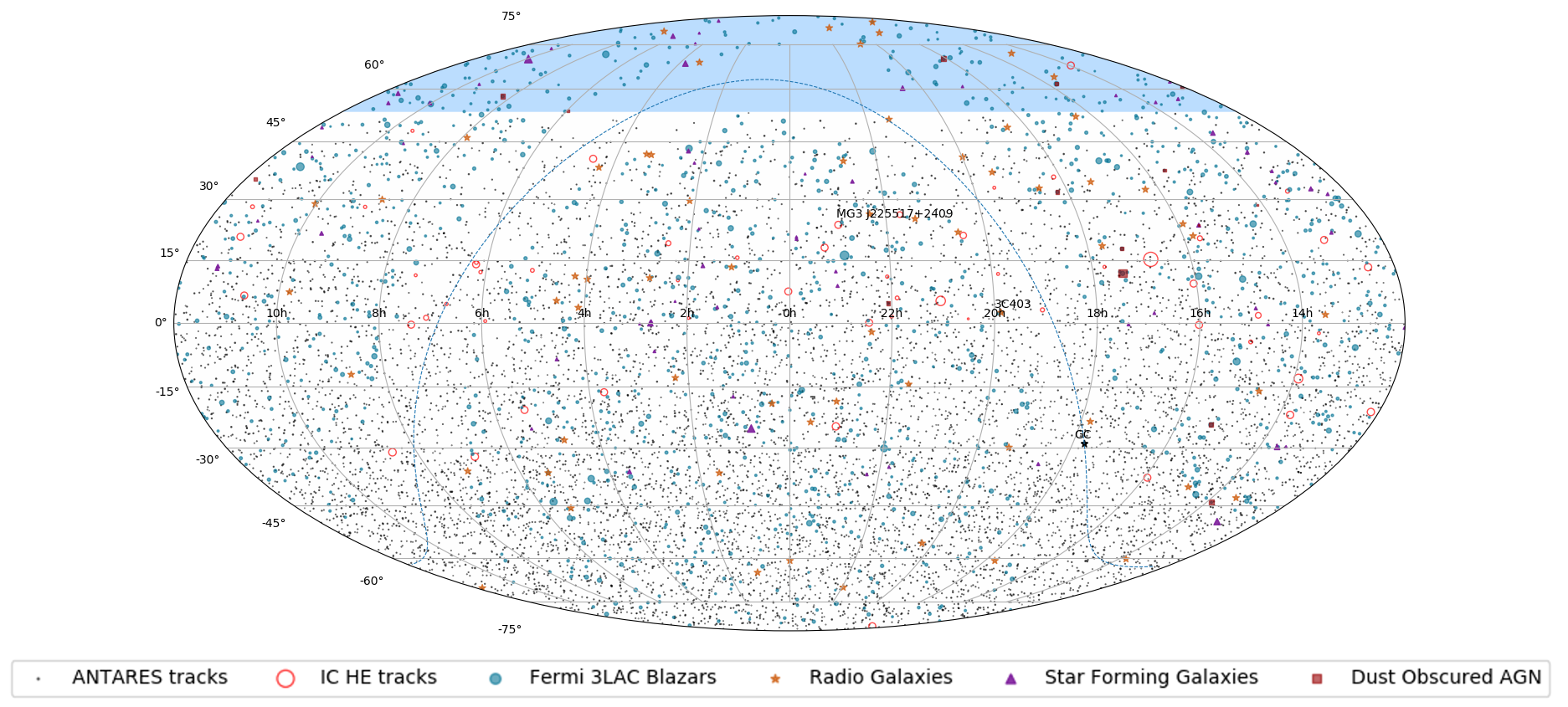}
\caption{Sky map in equatorial coordinates of the 8754 ANTARES track events (small black circles), together with the considered target sources: \textit{Fermi} 3LAC blazars (blue circles), radio galaxies (orange stars), star-forming galaxies (violet triangles), dust-obscured AGN (red squares), and IceCube tracks (red empty circles). For astrophysical sources, the radius of each marker is proportional to the square root of the weight used in the likelihood. For IceCube tracks, the radius of each circle is proportional to the angular error (displayed angular size not to scale on the figure for clarity). The Galactic Center is indicated by a black star and the Galactic Plane by a blue dashed line. The blazar MG3 J225517+2409 and the radio galaxy 3C403 have their position indicated by their name. The blue shaded area corresponds to declinations out of the field of view of ANTARES. }
\label{fig:full_skymap}
\end{figure*}
\section{\label{sec:level4}Target sources}

The different catalogs used in the analysis are presented in the following. Figure 3 shows the position of all the considered astrophysical objects on a sky map in equatorial coordinates, together with the position of the candidate neutrinos detected by ANTARES.

\subsection{\textit{Fermi} 3LAC Blazars}

The \textit{Fermi} 3LAC catalog \cite{Fermi3LAC} contains blazars detected in gamma-rays between $1-300$ GeV by the \textit{Fermi}-LAT in 4 years of operation. 
The previous release (\textit{Fermi} 2LAC) has been used by the IceCube Collaboration in a stacking analysis \cite{ICBlazars}, where several sub-classes of objects have been tested. A similar approach is followed here: in addition to the full \textit{Fermi} 3LAC catalog, two additional sub-samples are also considered, as described below.\\

Blazars have a typical spectral energy distribution (SED) of their electromagnetic spectrum, that presents two bumps: a low-energy peak between IR and UV-X due to synchrotron emission of electrons, and a high-energy peak in the hard-X-rays/gamma-ray band that can be produced by both hadronic or leptonic processes. 
Blazars are classified into flat spectrum radio quasars (FSRQs) and BL Lacs, according to the presence (FSRQs) or absence (BL Lacs) of optical broad emission lines on top of the continuum. When this classification is not possible, the blazars are labelled as blazar candidate of uncertain type (BCU) in the catalog. \\ 

In addition, a second classification can be made using the position of the synchrotron peak to distinguish between low synchrotron peak (LSP), intermediate and high synchrotron peak (ISP and HSP respectively). This classification, introduced in \cite{SED}, was based on the observation of a sequence in the blazar population \cite{Fossati}: the most luminous objects having their synchrotron peak at lower frequencies. 


In addition, the objects that are flagged as doubtful in the \textit{Fermi} catalog are removed from the analysis (see \cite{Fermi3LAC}, section 2.)
The different (sub)samples that could be used in the analysis are:
\begin{itemize}
 \item \textit{Fermi} 3LAC: 1420 objects, with 1255 in the Field of View (FoV) of ANTARES,
 \item FSRQ: 414 objects (382 in FoV),
 \item BL Lacs: 604 objects (509 in FoV),
 \item LSP: 666 objects (604 in FoV),
 \item ISP+HSP: 685 objects (590 in FoV).
\end{itemize}

The overlap between sub-samples is indicated in the following table:  

\begin{table*}[!h]
\begin{center}

\begin{tabular*}{0.5\textwidth}{@{\extracolsep{\fill}}|c||c|c|c|c|}
   \hline 
    & FSRQ & BL Lac & BCU & Total \\
   \hline
  LSP  & 365 & 148 & 153 & 666  \\
  \hline
  ISP + HSP &  44 & 440 & 201 & 685 \\
  \hline
 Unclassified  & 5 & 16 & 48 & 69  \\
  \hline
 Total &  414 & 604 & 402 & 1420   \\
 \hline
\end{tabular*}
\caption{Repartition of the \textit{Fermi} 3LAC objects into the different sub-samples.}
\end{center}
\end{table*}

As can be seen, most of the FSRQ are classified as LSP and the majority of BL Lacs are classified as ISP+HSP. 
In order to reduce the number of trials, the analysis is then restricted to the FSRQ and the BL Lac sub-samples only, in addition to the full \textit{Fermi} 3LAC sample.\\

For the implementation in the likelihood, the measured spectral slope $\gamma$ of the gamma-ray energy spectrum is taken into account in the weights. This is done by using the energy flux: $w^\mathrm{model} = \int E\cdot \Phi_0 \left(\frac{E}{E_0}\right)^{-\gamma} dE $ where the integral runs over the \textit{Fermi}-LAT energy range $[1 - 100 ] $ GeV.

\subsection{Dust-obscured AGN}

The authors of \cite{ObscuredAGN} suggest to consider a specific class of AGN where the jet axis is pointing towards the Earth, but passing through a high quantity of dust or gas, thus potentially enhancing the neutrino production rate via hadronic p-p interactions.

The sample is built from cross-correlating objects from a radio-emitting galaxies catalog \cite{VanVelzen} with the \textit{Fermi} 2LAC catalog \cite{Fermi2LAC}.
The morphologies that are incompatible with a jet viewed face-on are rejected. The final selection criteria uses the observed X-ray flux of the objects, that should be low compared to the radio flux because of the attenuation in the surrounding matter.

The 25\% weakest X-ray emitters are then selected, leading to a small catalog of 15 objects, containing 3~FSRQs, 1~ULIRG (Ultra Luminous IR Galaxy), and 11 BL Lacs.

To take into account the distance and the intrinsic luminosity, the radio flux density measured at $1.4$ GHz is used as a weight in the likelihood: $w^\mathrm{model} = \Phi_\mathrm{Radio}.$

\subsection{Radio galaxies}

Radio galaxies are a sub-class of active galaxies, that could be considered as possible sources of high-energy neutrinos \cite{Hooper_2016}. The neutrinos could be produced by the hadronic interaction of accelerated cosmic rays within the jets or in the giant lobes at the end of the jets. 

Radio galaxies have an active nucleus emitting jets of relativistic plasma, but contrary to blazars, the direction of the jet axis is not close to our line of sight. The radio galaxies are then in general less luminous than blazars individually, but they are more abundant in the local universe.

A sample of 65 soft gamma-ray selected radio galaxies suggested by L. Bassani is considered in this study. The authors of \cite{Radiogals} looked at extended radio galaxies in the sample of soft gamma-ray ($20-100$ keV) selected AGN in either the INTEGRAL or SWIFT-BAT surveys.
The selection procedure yields a sample that contains the brightest and most accretion-efficient radio galaxies in the local sky, with very powerful jets producing an extended double lobed radio morphology. 

Among the published list of objects, one source without redshift information is removed, and Centaurus A is excluded because it cannot be considered as a point-source (the angular extension of its radio lobes is $\sim10^\circ$).\\ 

The weight $w^\mathrm{model} = L_\gamma/d^2$ is used, where $L_\gamma$ is the luminosity measured in soft-gamma rays by SWIFT-BAT (except for five galaxies where the luminosity measured by INTEGRAL is used) and $d$ is the luminosity distance to the source. 
The luminosity distance is computed for each galaxy adopting the redshift indicated in \cite{Radiogals}, using a standard flat $\Lambda$CDM cosmology with parameters : 
$H_0 = 71$ km s$^{-1}$Mpc$^{-1}$, $\Omega_m=0.27$ and $\Omega_\Lambda =0.73$ \cite{WMAP9yr}.

\subsection{Star Forming Galaxies}

A catalog of 64 star forming galaxies (SFG) published by the \textit{Fermi}-LAT Collaboration \cite{SFGPaper} was used recently by the Pierre Auger Collaboration, reporting a $3.9 \, \sigma$ correlation with ultra high-energy cosmic rays (UHECRs) with energy above $39$ EeV \cite{Auger4sigma}.
The SFG do not have necessarily an active nucleus, but have a strong star formation rate traced by the intense infrared radiation. The high interstellar gas density can act as target material for the cosmic rays that are trapped in the galactic magnetic field, thus potentially producing neutrinos.

The dominant galaxies in terms of emitted infrared flux are: M82, NGC 253 and NGC 4945. Due to the selection procedure (detection in gamma-rays by \textit{Fermi}-LAT), most of the objects present in the catalog are very nearby: 5~objects are located within a radius of 5 Mpc, 14 galaxies are within 10 Mpc and 51 within 100 Mpc. A caveat of the use of this catalog is then its incompleteness due to the gamma-ray selection.

The following weights are used in the likelihood: $w^\mathrm{model} = L_\mathrm{IR}/d^2$ where $L_\mathrm{IR}$ is the total IR luminosity $[8-1000]$~$\mu \mathrm{m}$ and $d$ is the luminosity distance to the source.

\subsection{IceCube High-Energy tracks}

The high-energy tracks measured by IceCube are an interesting sample for a stacking analysis: they are likely to be of astrophysical origin and have for the majority an angular resolution of $\sim 1 ^\circ$. Their directions are then in general a sufficiently good proxy for the location of neutrino sources. 

The sample consists of 56 events, 55 in the FoV of ANTARES: 35 tracks from the IceCube 8 year sample of up-going muons  with deposited energy above $200$ TeV \cite{IC_Spectrum} and 21 additional tracks from the 6 year HESE sample \cite{IC_HESE6}. The majority of those events are located in the Northern hemisphere, and have been reconstructed with angular errors varying from $\sim 0.5^\circ$ up to $\sim 5^\circ$. A complementary study of the correlation between IceCube high-energy tracks and ANTARES point-source data is reported in  \cite{GiuliaPS}, where each track is considered separately. \\

\begin{table*}
\caption{Results of the likelihood stacking analysis.
The test statistics values are reported as $TS$, the pre-trial p-values are labelled as $p$, and the post-trial as $P$. The 90\% CL $\nu_{\mu} + \bar{\nu}_{\mu}$ flux upper limits are expressed in terms of the total $E^{-2}$ flux normalization at 1 GeV for the considered objects (in units of $10^{-8}$ GeV$^{-1}$cm$^{-2}$s$^{-1})$.}
\begin{ruledtabular}
\begin{tabular}{|c||c|c|c|c|c|c|c|c|}
    & \multicolumn{4}{c|}{Equal weighting} & \multicolumn{4}{c|}{Flux weighting} \\
   \hline
   Catalog & $TS$ & $p$ & $P$ & $\Phi_{90\%}^\mathrm{UL}$  & $TS$ & $p$ & $P$ & $\Phi_{90\%}^\mathrm{UL}$  \\
   \hline
   \hline
    \textit{Fermi} 3LAC All Blazars & 6.1 & 0.19 &  0.83  & 4.3  & 0.21 & 0.85 & 1.0 & 2.1\\
  \hline
  \textit{Fermi} 3LAC FSRQ & 0.83 & 0.57  & 0.97 & 2.2 & $\sim 0$ & $\sim 1$ & 1.0 & 1.8\\
  \hline
  \textit{Fermi} 3LAC BL Lacs & 8.3 & 0.088 & 0.64 & 4.8 &0.84 & 0.56 & 0.96 & 2.0 \\
  \hline
  Radio Galaxies & 3.4 & $4.8 \cdot 10^{-3}$ & 0.10 & 4.2 & 5.1 & $6.9 \cdot 10^{-3}$ & 0.13 & 4.7 \\
   \hline
  Star Forming Galaxies & 0.030 & 0.37 & 0.93 & 2.0  & $\sim 0$ & $\sim 1$  & 1.0 &1.7 \\
   \hline
  Dust-obscured AGN  & $1.0 \cdot 10^{-3}$ & 0.73 & 0.98 & 1.5     &$\sim 0$ &$\sim 1$  & 1.0 & 1.4 \\
   \hline
  IC HE Tracks & 0.77 & 0.05 & 0.49 & 5.2   & - & - & - & -\\
\end{tabular}
\end{ruledtabular}
\end{table*}

The IceCube candidate neutrinos have individual angular errors that need to be accounted for when searching for point-source excesses. The likelihood method is then modified as follows: an independent fit is performed for each IceCube track in the sample, where the location in $(RA,\delta)$ is let free to vary in the fit within a range of $\pm 2\sigma$ separately for each coordinate, where $\sigma$ is the angular error reported by IceCube (for the 8 year up-going muon sample, the $50\%$ containment value reported in the article is multiplied by a factor $1.5$).

The test statistic is then defined as the sum over all the individual test statistics:
\begin{equation}
TS=\frac{1}{\sum_j w_j} \sum_{j=1}^{N_\mathrm{sources}} w_j \cdot TS_j,
\end{equation}
where $TS_j$ is the local test statistic obtained for the $j^\mathrm{th}$ IceCube track, carrying a weight $w_j=\mathcal{A}(\delta_j)$ equal to the ANTARES acceptance at the corresponding declination. 

The global significance is evaluated via the total test statistic $TS$,  nevertheless, the relative contribution of each IceCube track can be examined by looking at the distribution of individual test statistics $TS_j$.\\

\section{\label{sec:level5}Results}

The results of the stacking analysis are summarized in table 2.  The trial factor computation has been performed by generating $10^4$ pseudo-experiments with only background events (randomized samples), and applying the same likelihood analysis for the 10 different combinations of catalogs and weighting schemes considered in this analysis. 

None of the tested catalogs shows a significant association with the 11 year point-source sample of neutrinos detected by ANTARES. The smallest p-value is obtained for the radio galaxy catalog with the equal weights hypothesis, with a pre-trial p-value $p=4.8 \cdot 10^{-3}$, equivalent to a $2.8 \, \sigma$ excess (two-sided convention), reduced to $1.6 \, \sigma$ post-trial. 

The 90\% CL flux upper limits in table 1 are expressed in terms of the one-flavor ($\nu_{\mu} + \bar{\nu}_{\mu}$) total flux normalization at 1 GeV:
\begin{equation}
\Phi^0_{\rm tot}=\sum_{j=1}^{N_{\rm sources}} \Phi^0_{j},
\end{equation}
computed for a spectral index $\gamma=2$.
For comparison, in  \cite{ICBlazars_ICRC19} the IceCube Collaboration reports the results of a similar stacking likelihood analysis using 1301 blazars contained in the \textit{Fermi} 3FHL catalog \cite{Fermi3FHL}. The quoted upper limit for an $E^{-2}$ spectrum is $\Phi^{\rm UL}_{90\%}=7.2 \cdot10^{-9}$ GeV$^{-1}$ cm$^{-2}$s$^{-1}$ which is a factor $\sim 6$ more stringent than the ANTARES limit presented here.


\subsection{\label{sec:subsec1}Search for dominant sources}
\begin{figure}[htbp]\centering
\includegraphics[width=0.49\linewidth]{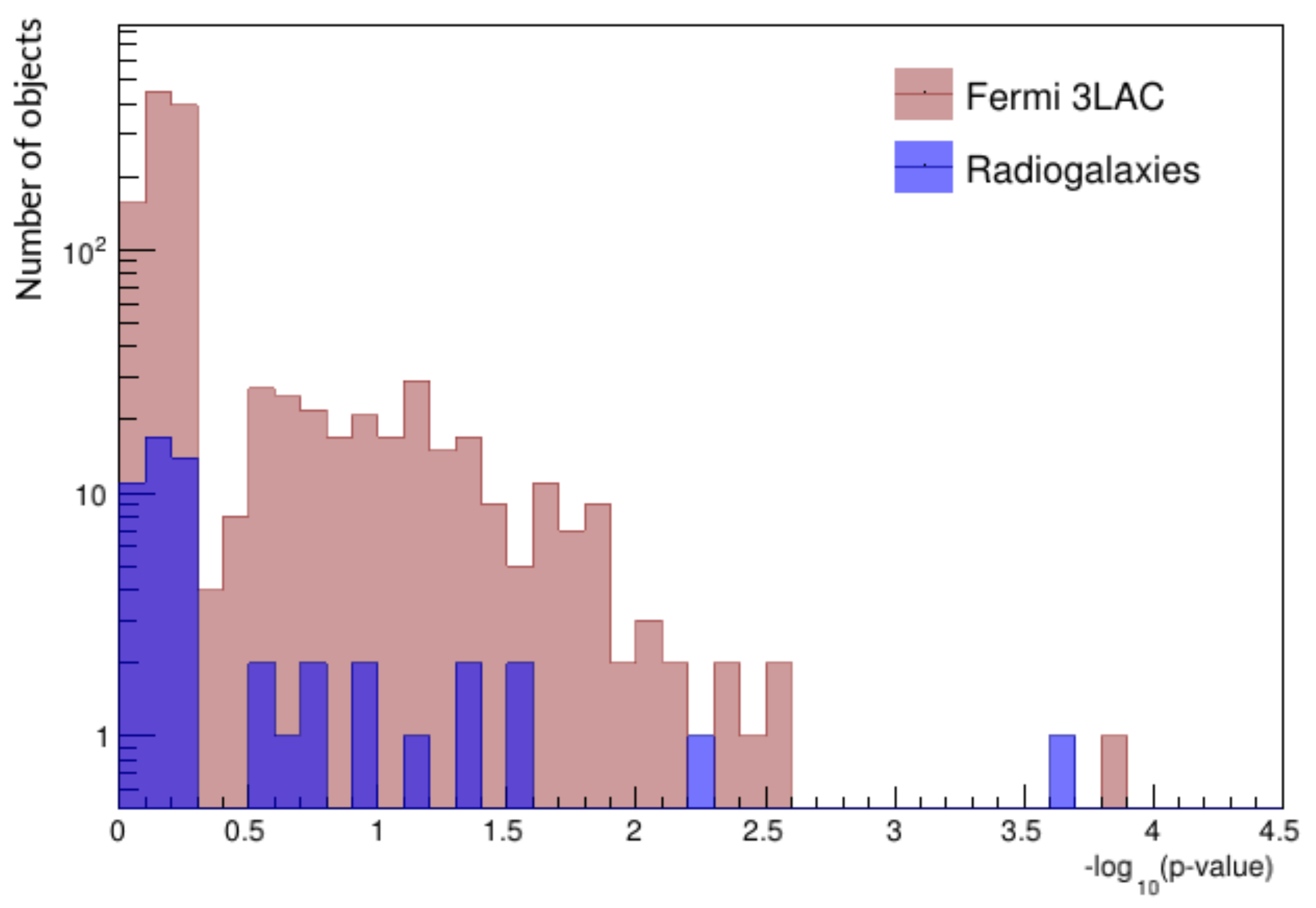}
\caption{Distribution of the individual p-values for the \textit{Fermi} 3LAC blazars and for the radio galaxies.}
\label{fig:pvals}
\end{figure}

To search for a possible dominant contribution of a small number of sources, an individual likelihood fit has been performed for each object in the radio galaxy and \textit{Fermi} 3LAC catalogs. The resulting p-values associated to each source are shown in figure \ref{fig:pvals}. Apart from the bulk of the distribution, a single source is present in the tail of the distribution of p-values for both the radio galaxies and the \textit{Fermi} blazars. Their properties are thus investigated in more details in the following sections.\\ 

\begin{figure*}[t]\centering
\includegraphics[width=0.49\linewidth]{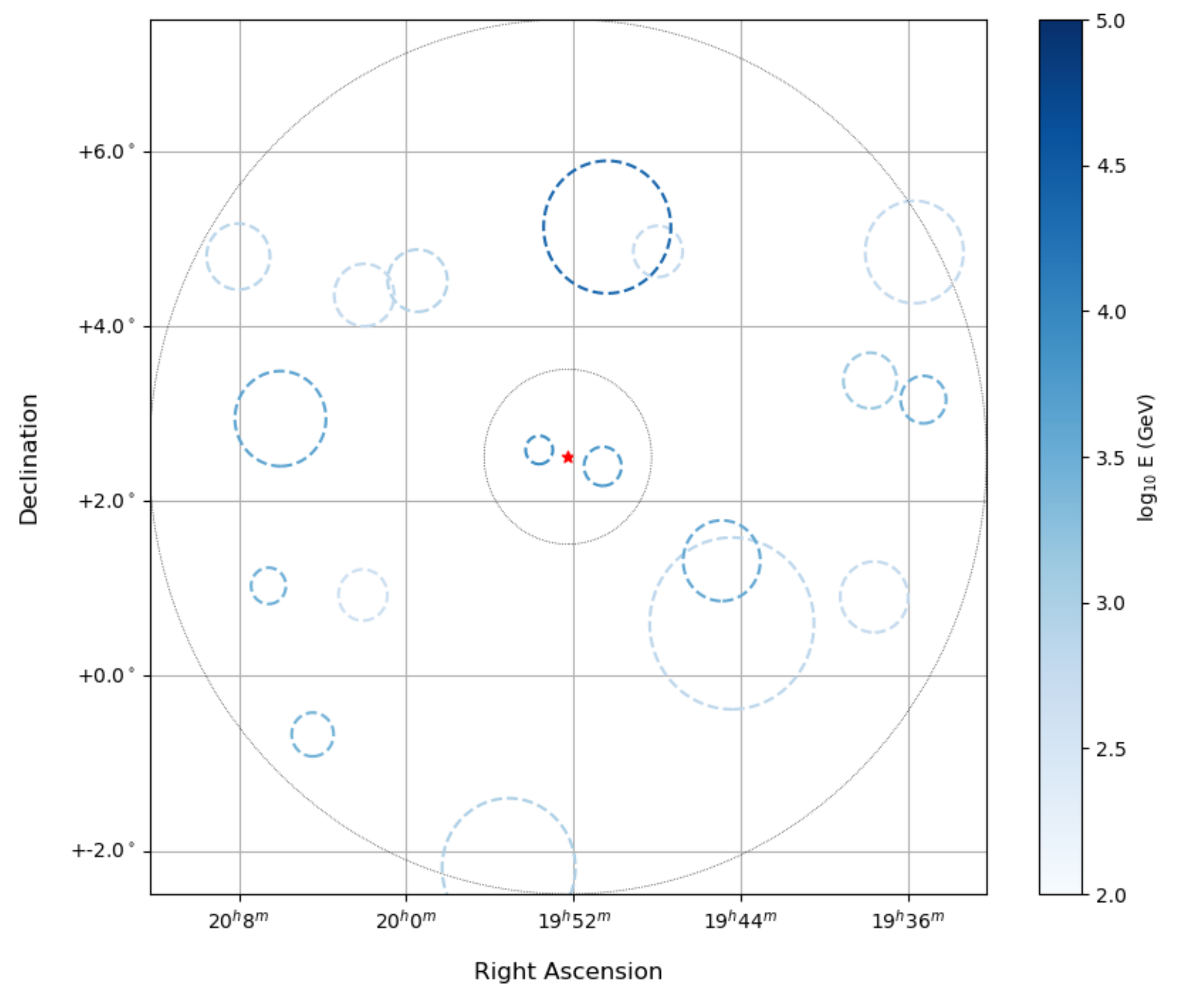}
\hfill
\includegraphics[width=0.49\linewidth]{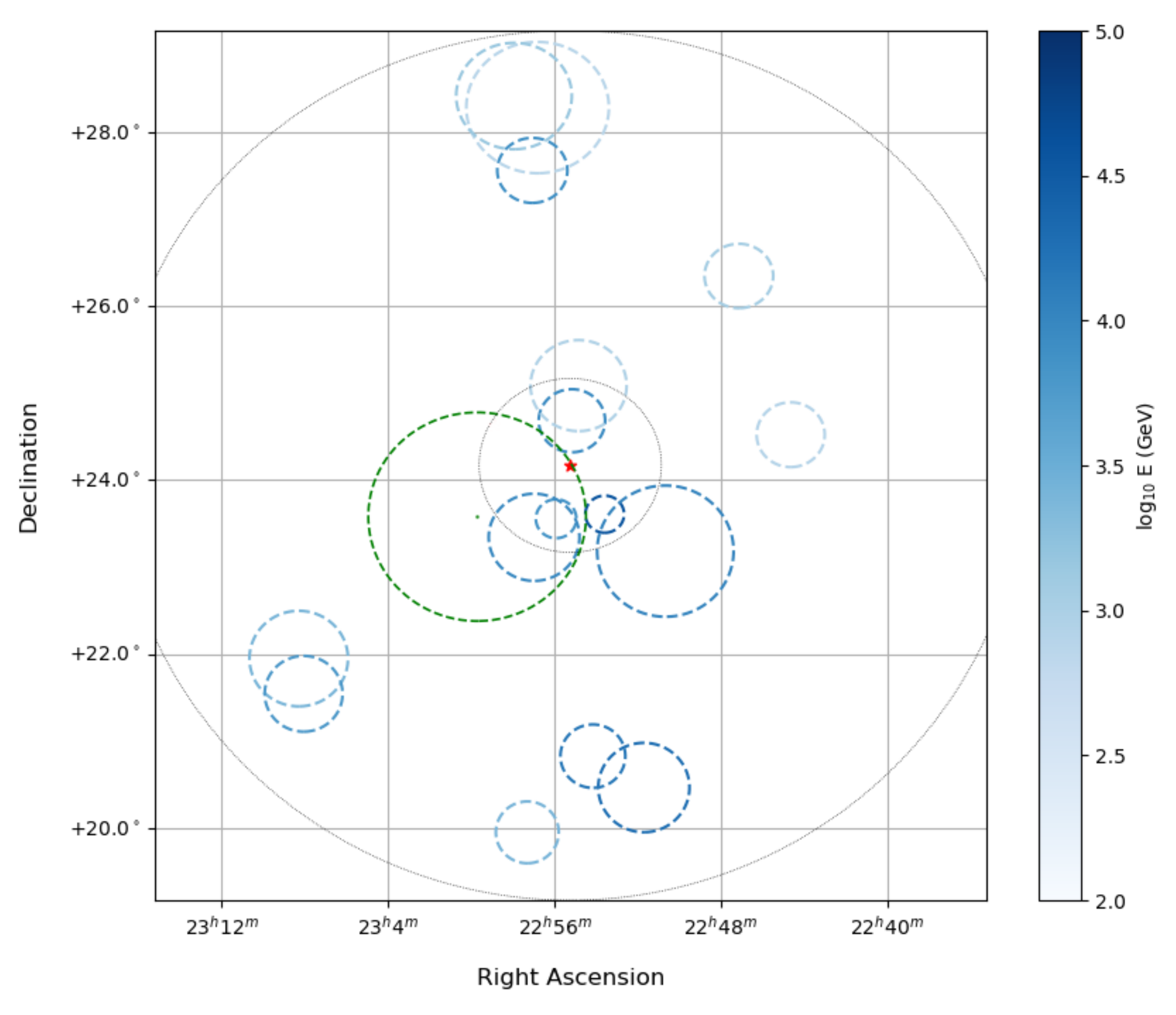}
\caption{Arrival directions of ANTARES track events in equatorial coordinates around the most prominent sources: \textit{left}: radio galaxy 3C403, \textit{right}: blazar MG3 J225517+2409. Each event is represented by a dashed empty circle with a radius equal to its angular error estimate. The estimated energy of each event is indicated by a color code. The thin black circles centered around the source (red star marker) indicate the $1^\circ$ and $5^\circ$ angular distances to the source. On the right plot, the IceCube track IC-100608A is indicated as a green dashed circle, with radius equal to the $50\%$ containment value ($\sim 1.2^\circ$).}
\label{fig:skyplots}
\end{figure*}

\paragraph{Radio galaxies} The radio galaxy with the smallest p-value is 3C403, which is found with a pre-trial p-value of $p=2.3 \cdot 10^{-4}$, equivalent to $3.7 \, \sigma$. The probability to find by chance a similar excess among the $N=56$ radio galaxies in the field of view of ANTARES has been computed using $P(n \geq 1)=1-(1-p)^N$, leading to a value $P=1.3\cdot 10^{-2}$, equivalent to $2.5 \, \sigma$.\\ 

The arrival direction of ANTARES events around this source are shown in figure \ref{fig:skyplots} (left), where one can see that the excess comes from the presence of 2 events lying at less than $0.5^\circ$ from the source. Those track events have an estimated angular uncertainty of $0.2^\circ$, and a reconstructed energy of $\sim5$ TeV and $\sim10$ TeV, respectively.\\  
The radio galaxy is located at a declination of $+2.5^\circ$ and its distance is evaluated to be $\sim$ 260 Mpc. The galaxy 3C403 is an FRII that presents a peculiar radio morphology, called ``-X-shape'' or ``-Winged" (see figure 1 of  \cite{XShape}) due to a recent change of the jet axis orientation, that could be the signature of a binary black hole merger or the fusion with a small galaxy. The interaction of the newly formed jet with the radiation and matter environment around the black hole could lead to high-energy neutrino emission, as suggested in  \cite{KunMerger}.

\paragraph{Blazars}
The most significant blazar from the \textit{Fermi} 3LAC catalog is the BL Lac MG3 J225517+2409 (\textit{Fermi} source 3FGL J2255.1+2411, or 4FGL J2255.2+2411 in the latest version of the catalog). A recent compilation of multi-wavelength observations for this object is reported in figure 8 of  \cite{DissectBlazars}, showing a synchrotron peak around $10^{15}$ Hz, indicating that the blazar can be classified in the ISP category, similar to TXS 0506+056. The redshift of MG3 J225517+2409 is currently unknown, but a limit of $z>0.86$ has been recently reported in  \cite{redshiftMG3}.  

The gamma-ray flux of MG3 J225517+2409 measured by \textit{Fermi}-LAT in 8 years of observation \cite{Fermi4FGL} is well described by a simple power law without cut-off, with a spectral index $\gamma = 2.10 \pm 0.05$, with an integral flux of $\Phi=1.0 \cdot 10^{-9}$ photons cm$^{-2}$s$^{-1}$ for energies between $1$ and $100$ GeV.

It is noteworthy to remark that another candidate blazar 3FGL J2258.8+2437 is located only $\sim 1^\circ$ away from MG3~J225517+2409, and $\sim 1.2^\circ$ away from the position of the IceCube high-energy track event IC-100608A. However, this source does not have an identified counterpart and is not present in the 10 year \textit{Fermi}-LAT catalogs, suggesting that 3FGL J2258.8+2437 might have been a spurious detection. For the present study, this source is thus not considered in the chance coincidence calculation.


The result of the individual likelihood fit for this source gives a pre-trial p-value of $p=1.4 \cdot 10^{-4}$, equivalent to $3.8 \, \sigma$. The probability to find by chance a similar excess among the $N=1255$ blazars in the field of view of ANTARES is evaluated by generating pseudo-experiments, giving a post-trial value $P=0.15$ (equivalent to $1.4 \, \sigma$). The simple binomial evaluation gives a very similar value, $P(n \geq 1)=1-(1-p)^N \simeq 0.16$, as the overlap between sources is very small: there are only $58$ pairs of 3LAC objects that are separated by an angular distance less than $1^\circ$.\\

The distribution of ANTARES events around the blazar is shown in figure \ref{fig:skyplots} (right). There are 5 ANTARES tracks lying at less than $1^\circ$ from the source, with estimated energies ranging from $\sim3$ to $\sim 40$ TeV, and with estimated angular uncertainty between $0.2^\circ$ and $0.5^\circ$. This particular region of the sky has been previously reported by the ANTARES Collaboration \cite{GiuliaPS} to contain the hottest spot of the full sky point-source blind search; the maximum excess lies at $<0.7^\circ$ from the position of the blazar MG3~J225517+2409. 

Under the (restrictive) assumption that a neutrino flux is produced steadily by the source with an  $E^{-2}$ spectrum, an estimation of the flux normalization at 1 GeV of MG3 J225517+2409 with the ANTARES data would be $\Phi_0 \sim (0.6-3.0) \cdot 10^{-8}$ GeV$^{-1}$cm$^{-2}$s$^{-1}$, depending on the number of neutrino candidates that are considered as signal (between 1 and 5) from this source.\\

In addition to the ANTARES measurement, a high-energy muon track IC-100608A from the IceCube 8 year up-going muon sample \cite{IC_Spectrum} is found to lie at an angular distance of $1.1^\circ$ from the source. This event has a deposited energy in the IceCube detector of about $\sim 300 $ TeV, but it has a poor angular resolution: the $50\%$ confidence level containment region has a radius $\sim 1^\circ$, but the estimated $90\%$ CL region has a radius larger than $\sim 3^\circ$.
This poor angular reconstruction explains why the spatial correlation with MG3 J225517+2409 has not been reported before, the IceCube muon event IC-100608A being excluded from the selection in previous blazar correlation searches \cite{Padovani2016}, \cite{DissectBlazars}. 

\subsection{\label{sec:subsec2}Dedicated analysis of the region around the blazar MG3 J225517+2409}

\begin{figure}[htbp]\centering
\includegraphics[width=0.69\linewidth]{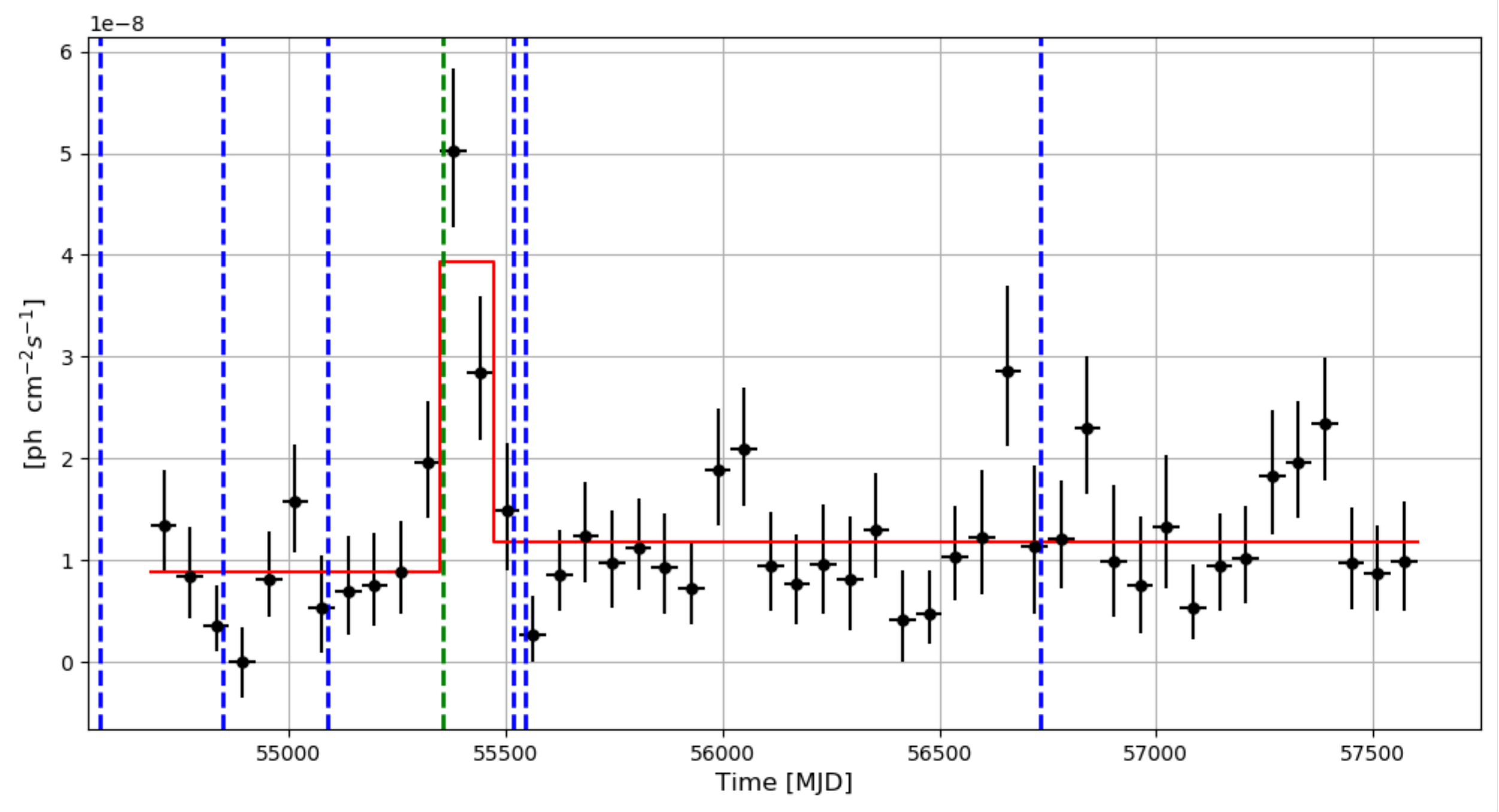}
\caption{\textit{Fermi}-LAT measured gamma-ray flux for the BL Lac 4FGL J2255.2+2411, as a function of time in Modified Julian Date (black points). The red line represents the model gamma-ray light curve that is used in the time dependent likelihood (obtained via a Bayesian block algorithm). The blue vertical lines indicate the arrival time of ANTARES events located at an angular distance smaller than $2^\circ$ from the blazar. The vertical green line corresponds to the arrival time of the IceCube track IC-100608A.}
\label{fig:lightcurve}
\end{figure}

The possible association between ANTARES and IceCube neutrinos and the blazar MG3 J225517+2409 is investigated in more detail in this section, by further looking at the time evolution of the gamma-ray emission of the source. The latest data from the \textit{Fermi} 4FGL catalog \cite{Fermi4FGL} is used to compare the evolution of the gamma-ray flux with the arrival times of ANTARES and IceCube events.

The \textit{Fermi} 4FGL catalog provides measurements with twice as much exposure as the \textit{Fermi} 3FGL, with $20\%$ larger acceptance, improved angular resolution above 3 GeV, a better model of the background galactic diffuse emission and a refined likelihood analysis. The 4FGL catalog was not available at the time of the stacking analysis, it is used only in this section for the time analysis.

The time evolution of the gamma-ray flux measured by \textit{Fermi}-LAT for the blazar 4FGL J2255.2+2411 above 100~MeV is shown in figure \ref{fig:lightcurve}, along with the arrival times of spatially coincident IceCube and ANTARES neutrino events. The data are presented in 48 bins of two-months width, from August 2008 to August 2016. A Bayesian block algorithm \cite{BayesianB} has been applied to the flux data in order to pinpoint possible flaring states. A significant increase of the flux by a factor of $\sim 4$ with respect to the average value is evident in a time window of $\sim 4$ months, centered at about MJD $= 55400$.


\subsubsection{ANTARES events}
The arrival times of the 5 ANTARES events around the blazar MG3 J225517+2409 are clustered at earlier observation times, in a large time window encompassing the flare but with none of them actually occurring during the high gamma-ray activity period (the two latest events were observed $\sim2$ and $\sim3$ months after the end of the gamma-ray flare). Therefore, including a time-dependent term in the likelihood analysis does not strengthen the potential association between ANTARES candidate neutrino events and the gamma-ray emission of the blazar. \\

Considering only the time information, a very simple and model independent test consists of comparing the average time distance to the flare center ($t_F$): \begin{equation}\tau=\frac{1}{N} \, \sum_{i=1}^{N=5} | t_i - t_F|,\end{equation}observed for the 5 ANTARES events, to the distribution that is expected from random pseudo-experiments (following the true time distribution of ANTARES events). 

The value obtained for the data is $\tau\simeq 400$ days, and the fraction of pseudo-experiments having a lower or equal value is found to be $\simeq2\%$, thus leading to an estimated p-value of $p=0.02$ ($2.3 \, \sigma$). \\

\begin{figure}[!h]\centering
\includegraphics[width=0.5\linewidth]{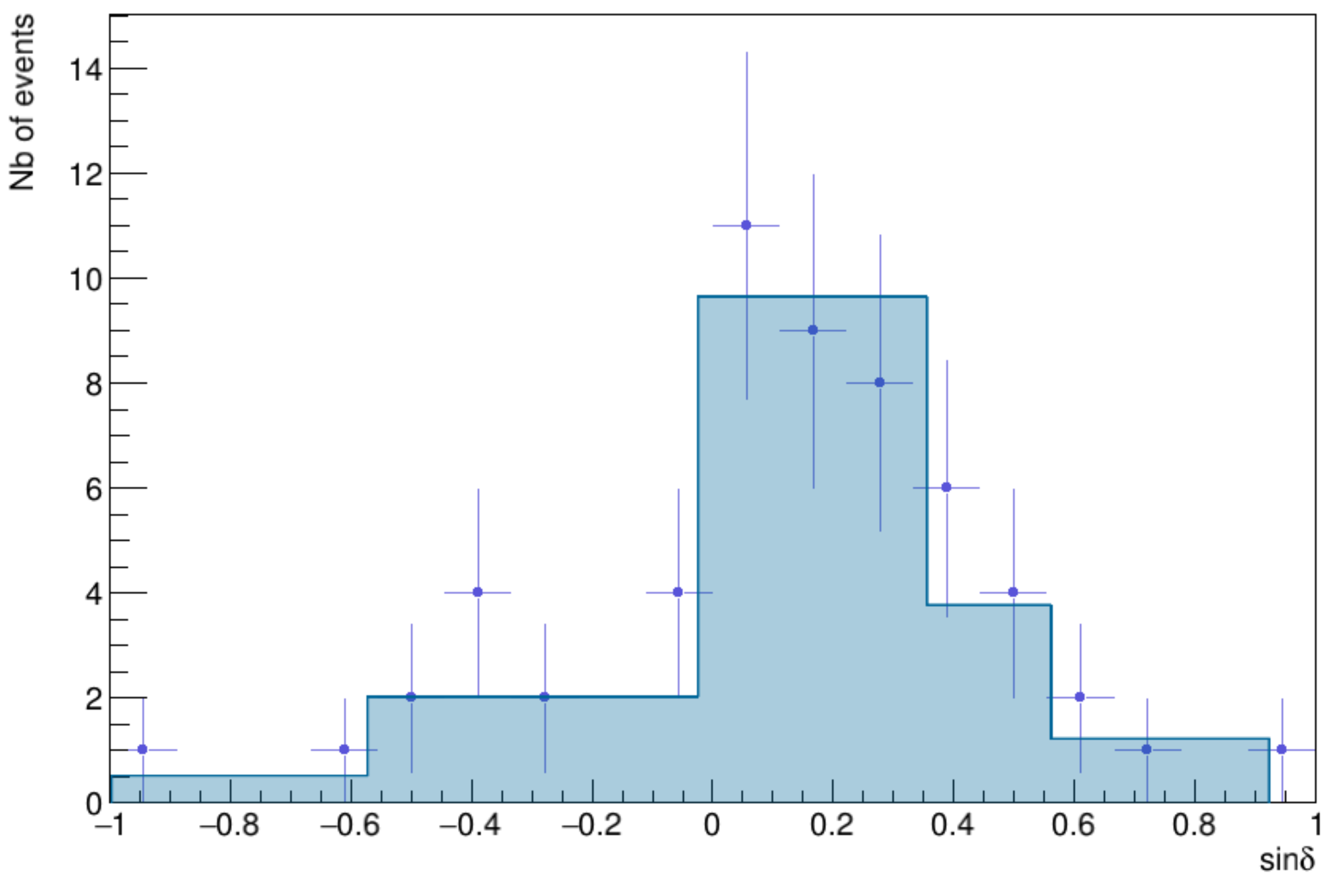}
\caption{Distribution of the sinus of the declination of the 56 high-energy track-like events detected by IceCube considered in this study. The blue filled histogram represents the Bayesian block fit to the data that is used in the likelihood analysis.}
\label{fig:IC_pdfs}
\end{figure}

\subsubsection{IceCube event}

To evaluate the degree of space and time correlation between the gamma-ray flux of MG3 J225517+2409 and the neutrino candidate of IceCube, an additional likelihood analysis has been performed for this specific event.

The likelihood of equation 1 is modified as follows:
\begin{equation} \ln \mathcal{L}(x | \mathrm{H}_1 ) = \ln \left[ \mu_s S(x_i)f_S(t_i) + \mu_b B(x_i)f_B(t_i) \right]  - \mu_s - \mu_b ,
\end{equation}
where the new terms $f_S(t_i)$ and $f_B(t_i)$ represent the signal and background time PDFs respectively. 

The signal $f_S(t)$ is chosen to be proportional to the Bayesian block estimation \cite{BayesianB} of the gamma-ray flux of MG3 J225517+2409, shown as a solid red curve in figure \ref{fig:lightcurve}. The background $f_B(t)$ is proportional to the average number of events detected per day by IceCube, using the point-source selection cuts \cite{IC8yrPS}. The value of $f_B(t)$ is assumed to be constant during the different data taking periods, from the IC59 sample (2009-2010) where  $f_B(t) \simeq 60$ events/day, up to the IC86 sample where $f_B(t) \simeq 210$  events/day.\\

The spatial PDF $S(x)$ is taken as a two-dimensional Gaussian function of width $\sigma=1.8 ^\circ$, corresponding to the  $90\%$ CL statistical angular error of $\sim3^\circ$ reported in  \cite{IC_Spectrum}. The background distribution $B(x)$ is assumed to be uniform in right ascension, and proportional to the value of a Bayesian block fit of the distribution in declination of the 56 IceCube high-energy tracks considered in the stacking analysis, as shown in figure \ref{fig:IC_pdfs}.\\

The result of the likelihood fit gives a fitted number of signal events $n_s=0.68$ and a test statistic equal to $TS=0.39$. The associated p-value is computed by generating pseudo-experiments, leading to a value $p=5.3 \cdot  10^{-2}$, equivalent to $1.9 \, \sigma$.  As the considered source has been specified \emph{a priori} using the ANTARES result, there are no trial factors to account for.\\
 
\section{\label{sec:level6}Discussion and conclusion}
The results of a stacking point-source likelihood search using 11 years of ANTARES data using muon neutrino candidates reconstructed as tracks have been presented. After accounting for trial factors, none of the catalogs considered have shown a significant result in the stacking analysis.

The most significant value is obtained for the radio galaxy catalog, with a pre-trial value of $p=4.8\cdot 10^{-3}$, reducing to $P=0.1$ post-trial. Of the individual sources, the radio galaxy 3C403 ($P=0.013$) and the blazar MG3~J225517+2409 ($P=0.15$) are reported as interesting targets.

For the \textit{Fermi} 3LAC blazars, the flux upper limits obtained in this study for an $E^{-2}$ neutrino spectrum and equal weights are a factor $\sim 6$ less stringent than the comparable IceCube limits presented in \cite{ICBlazars_ICRC19}. This also limits the maximum contribution of blazars to the IceCube astrophysical diffuse flux. Following the same procedure as \cite{ICBlazars_ICRC19}, which quotes a maximum 13.0--16.7\% contribution, our less-stringent limits yield a maximum contribution of $\sim80$--100\% at 90\% CL. Since ANTARES probes a lower energy range than that at which the IceCube flux is detected, we expect our limit to improve for steeper spectral indices, which are currently favoured by model fits \cite{ICAstroFluxLatest}.


For the radio galaxies, the mild excess found in ANTARES data gives flux upper limits of the same order of magnitude than for blazars, allowing for a non negligible contribution of this class of sources to the IceCube diffuse astrophysical flux. \\ 

In addition, a high-energy IceCube track is observed to be in coincidence with the blazar MG3~J225517+2409. This motivated an \emph{a posteriori} analysis of the space and time correlation between the gamma-ray flux measured by \textit{Fermi} and the neutrinos detected by ANTARES and IceCube for this source. The IceCube event is observed to be temporally coincident with what could be called a flaring period of the blazar. The chance probability of this association is estimated to be $p=5.3 \cdot  10^{-2}$ ($1.9 \, \sigma$). 

The five ANTARES events lying within $1^\circ$ from the source are not coincident in time with the flare. However, they all arrive within 400 days of the flare, with associated p-value of $p=0.02$. 
This result hardly reconciles a direct association between a TeV neutrino excess and MeV-GeV gamma-ray flare.
A similar finding was already highlighted for the apparent lack of association between the 2014/15 neutrino excess observed by IceCube from TXS~0506+056 and \textit{Fermi}-LAT activity observation \cite{TXSFollow,TwoFermiLatBlazars}.

The lack of gamma-rays in coincidence with the ANTARES neutrino events could be due to a high optical depth to photons at \textit{Fermi}-LAT energies. Given the current range of models for neutrino production in TXS~0506+056 \citep{Keivani2018_TXS, Reimer2019_TXS,Cao2020_TXS,Sahu2020_TXS,Petropoulou2020_TXS, Murase2018_TXS}, such a scenario seems plausible.

Combining the space and time probabilities for the ANTARES events using

$$
  p_\mathrm{space-time}=p_\mathrm{spatial} \cdot p_\mathrm{time} \, \left( 1-\ln (p_\mathrm{spatial} \cdot p_\mathrm{time}) \right),$$
leads to a value $ p_\mathrm{space-time}=  1.5 \cdot 10^{-2} \; (2.3 \, \sigma)$.\\
 Combining the ANTARES space-time and the IceCube p-values gives
\begin{eqnarray*}
p_\mathrm{combined}&&=p_\mathrm{ANT} \cdot p_\mathrm{IC} \, \left(1-\ln (p_\mathrm{ANT} \cdot p_\mathrm{IC}) \right) \\
&&= 8.0 \cdot 10^{-3} \; (2.6\, \sigma).
\end{eqnarray*}
Thus there is no strong statistical evidence for an association. Furthermore, we emphasize that these tests are done \emph{a posteriori}, and that the quoted p-values reported here are thus subject to bias. Nonetheless, given the suggestions of neutrino associations with other blazars \citep{Kadler2016_PeV_blazar,Investig2Blazars}, we advocate to include MG3~J225517+2409 as a potential interesting neutrino source for future studies.

The accumulation of independent neutrino data with larger statistics and better energy and angle resolution in the near future by next generation neutrino telescopes such as KM3NeT \cite{KM3NeT} will be of great interest to study the correlation between gamma-ray and neutrino emission. 

\acknowledgments
The authors thank Loredana Bassani for the suggestion of the radio galaxy catalog and for fruitful discussions.

The authors acknowledge the financial support of the funding agencies: Centre National de la Recherche Scientifique (CNRS), Commissariat à l’énergie atomique et aux énergies alternatives (CEA), Commission Européenne (FEDER fund and Marie Curie Program), Institut Universitaire de France (IUF), LabEx UnivEarthS (ANR-10-LABX-0023 and ANR-18-IDEX-0001), Région Île-de-France (DIM-ACAV), Région Alsace (contrat CPER), Région Provence-Alpes-Côte d’Azur, Département du Var and Ville de La Seyne-sur-Mer, France; Bundesministerium für Bildung und Forschung (BMBF), Germany; Istituto Nazionale di Fisica Nucleare (INFN), Italy; Nederlandse organisatie voor Wetenschappelijk Onderzoek (NWO), the Netherlands; Council of the President of the Russian Federation for young scientists and leading scientific schools supporting grants, Russia; Executive Unit for Financing Higher Education, Research, Development and Innovation (UEFISCDI), Romania; Ministerio de Ciencia, Innovación, Investigación y Universidades (MCIU): Programa Estatal de Generación de Conocimiento (refs. PGC2018-096663-B-C41, -A-C42, -B-C43, -B-C44) (MCIU/FEDER), Severo Ochoa Centre of Excellence and MultiDark Consolider (MCIU), Junta de Andalucía (ref. SOMM17/6104/UGR and A-FQM-053-UGR18), Generalitat Valenciana: Grisolía (ref. GRISOLIA/2018/119) and GenT (ref. CIDEGENT/2018/034), Spain; Ministry of Higher Education, Scientific Research and Professional Training, Morocco. We also acknowledge the technical support of Ifremer, AIM and Foselev Marine for the sea operation and the CC-IN2P3 for the computing facilities.

\bibliography{biblio}{}
\bibliographystyle{aasjournal}



\end{document}